# Spatial and doping effects on radiative recombination in thin-film near-field photonic energy converters

Dudong Feng, Shannon K. Yee, and Zhuomin M. Zhang*

George W. Woodruff School of Mechanical Engineering
Georgia Institute of Technology, Atlanta, GA 30332, USA

*Corresponding author: zhuomin.zhang@me.gatech.edu

**ABSTRACT:**

Modeling radiative recombination is crucial to the analysis of photonic energy converters. In this work, a local radiative recombination coefficient is defined and derived based on fluctuational electrodynamics that is applicable to thin-film cells in both the near field and far field. The predicted radiative recombination coefficient of an InAs cell deviates from the van Roosbroeck-Shockley relation when the thickness is less than 10 μm and the difference exceeds fourfold with a 10 nm film. The local radiative recombination coefficient is orders of magnitude higher when an InAs cell is configured in the near field. The local radiative recombination coefficient reduces as the doping level approaches that of a degenerate semiconductor. The maximum output power and efficiency of a thermoradiative cell would be apparently overpredicted if the luminescence coefficient (defined in this letter) were taken as unity for heavily doped semiconductors.

**KEYWORDS:** External luminescence, fluctuational electrodynamics, near-field radiative energy converter, radiative recombination coefficient

With the advantages of compact size and solid-state operation, photonic energy converters are promising technologies for future power generation or cooling applications.[1-4] Thermophotovoltaic (TPV) devices generate electrical current by absorbing photons from a hotter emitter, while thermoradiative (TR) cells generate electrical current by emitting photons to a colder object or surroundings.[5-7] The phenomenon of electroluminescence that enables light-emitting diodes can also be used to transfer thermal energy from a colder region to a hotter region while consuming electrical power.[9,10]

Near-field radiative heat transfer can boost the photon flux by orders of magnitude as demonstrated both theoretically and experimentally.[11,12] The enhancement of photon flux by operating in the near-field regime can boost the throughput as well as conversion efficiency with carefully designed structures, including plasmonic couplers and thin-film cells.[5-10] For heavily doped semiconductors, the effect of free carriers on the dielectric function needs to be taken into consideration in the modeling.[13-16] A number of groups have recently experimentally demonstrated near-field photonic energy converters.[17-21]

The annihilation of electron-hole pairs through the radiative recombination process is responsible for the emission of photons or luminescence. Quantitatively analyzing radiative recombination rate or coefficient is of critical importance to the evaluation of photonic energy converters including the dark current estimation.[22-24] Trupke et al.[25] studied the temperature dependence of radiative recombination coefficient using the van Roosbroeck-Shockley relation[26] considering the radiative properties of semi-transparent laminate samples. While external luminescence and photon recycling have been investigated for both far-field and near-field photonic energy converters,[27-30] the local radiative recombination coefficient is not clearly defined, especially in the near-field regime. Furthermore, the magnitude and impact of above-bandgap thermal radiation due to free carriers in heavily doped semiconductors on the device performance are yet to be explored.

In this work, fluctuational electrodynamics (FE) is used to connect the external luminescent emission to a local radiative recombination coefficient. A thin-film InAs cell is modeled in free space as well as in a thermophotovoltaic (TPV) setup with or without a back gapped reflector (BGR). Comparison is made with the conventional van Roosbroeck-Shockley relation for a free-standing film with varying thicknesses. The effect of doping for cells with high dopant concentrations is considered using a luminescence coefficient, which also enables the distinction between thermal and nonthermal radiation above the bandgap energy. The effect of luminescence coefficient on the performance of a TR device is quantitatively examined.

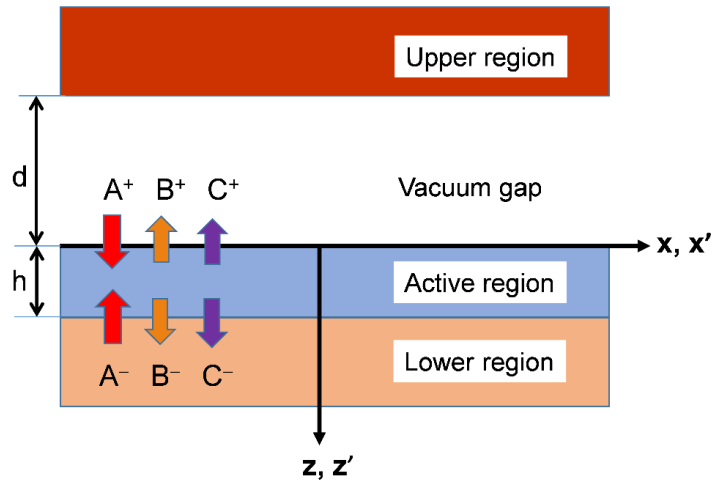

Fig. 1. Schematic of a thin-film radiative energy converter with four regions: an upper semi-infinite region, vacuum gap region with a thickness of $d$, active (cell) region with a thickness of $h$, and a lower semi-infinite region. The processes A, B, and C describe irradiation into the cell, thermal emission from the cell, and external electroluminescent emission from the cell across its upper (+) and lower (−) boundaries, respectively.

Consider a near-field photonic energy converter as shown in Fig. 1 with an active region (or cell) $0 \leq z' \leq h$. The cell receives incoming thermal radiation from the upper and lower regions (denoted by $A^+$, $A^-$). It also emits photons through thermal radiation ($B^+$, $B^-$) and luminescent emission ($C^+$, $C^-$). Depending on the temperature and voltage biases, Fig. 1 may describe a power generator or refrigerator.[1] Fluctuational electrodynamics establishes the relation between the resulting electromagnetic field with the randomly fluctuating charges at thermal equilibrium. For a source volume in a planar layer or film, the spectral radiative heat flux is evaluated from the Poynting vector at location $z$ and expressed as follows:[11,31]

$$q_\omega(\omega,z) = \frac{2\omega^2}{\pi c^2} \text{Re}\left[ i \int_{V'} \sum_{\alpha=x,y,z} \left( G^E_{x\alpha} G^{H*}_{y\alpha} - G^E_{y\alpha} G^{H*}_{x\alpha} \right) W(\omega,z') dV' \right] \quad (1)$$

Here, $\omega$ is the angular frequency, $c$ is the speed of light in vacuum, $i = \sqrt{-1}$ is the imaginary unit, $V'$ is the source volume, $G^{E,H}_{m\alpha}$ ($m = x, y$; $\alpha = x, y, z$) are the tensor components of the electric ($E$) or magnetic ($H$) dyadic Green's functions, and * denotes complex conjugate. The integration over the $x'-y'$ plane is carried out by transforming the Green's function to the wavevector $k_x - k_y$ space. For a multilayer structure, the spatial integration is with respect to $z'$ only. For thermal emission at equilibrium, $W(\omega,z') = \varepsilon''(\omega)\Theta(\omega,T)$, where $\varepsilon''$ is the imaginary part of the dielectric function at $z'$ and $\Theta(\omega,T) = \hbar\omega[\exp(\hbar\omega/k_B T)-1]^{-1}$ is the mean energy of a Planck oscillator at temperature $T$. Note that $k_B$ and $\hbar$ are the Boltzmann constant and the reduced Planck constant, respectively. Considering a direct bandgap semiconductor, the emission may be devided into a nonthermal portion due to interband transition that is affected by the chemical potential and a thermal portion due to other transitions such as free-carrier transitions and lattice vibrations. Therefore,

$$W(\omega,z') = \varepsilon_{ib}''(\omega)\Psi(\omega,T,\mu) + \left[\varepsilon_{fc}''(\omega) + \varepsilon_{la}''(\omega)\right]\Theta(\omega,T) \qquad (2)$$

where $\Psi(\omega,T,\mu) = \hbar\omega\{\exp[(\hbar\omega-\mu)/k_BT]-1\}^{-1}$ may be viewed as the mean energy of a photon mode based on the modified Bose-Einstein distribution considering photon chemical potential $\mu$.[2,32] Subscripts ib, fc, and la in Eq. (2) signify the contributions to the imaginary part of the dielectric function due to the interband transition, free carriers, and lattice vibrations, respectively, such that $\varepsilon'' = \varepsilon_{ib}'' + \varepsilon_{fc}'' + \varepsilon_{la}''$. The photon chemical potential cannot exceed the bandgap energy $E_g = \hbar\omega_g$. Furthermore, $\varepsilon_{ib}'' = 0$ when $\omega < \omega_g$. The significance of Eq. (2) is that it separates the luminescent emission originated from interband transition from the thermal emission at above bandgap energies.

The ratio of $\varepsilon_{ib}''$ and $\varepsilon''$ is proportional to the luminescence emission over the total radiative emission:

$$\phi(\omega) = \frac{\varepsilon_{ib}''(\omega)}{\varepsilon''(\omega)} \qquad (3)$$

Djuric et al.[33] used the ratio of absorption coeffients in a similar way to analyze the quantum efficiency of InSb photodiodes. The quantity defined in Eq. (3) was called the cell internal quantum efficiency.[13] In the present study, $\phi$ is termed luminescence coefficient since it is a factor that should be included in evaluating luminescent emission. For lightly doped semiconductors, $\phi$ is close to unity; however, the effect on $\phi$ must be considered for heavily doped semiconductors. The dielectric function model of InAs considering the effects of temperature and doping level is outline in the supplemantory material (SM) after Milovich et al.[16]

The photon flux due to external luminescence from the cell is calculated by adding the processes $C^+$ and $C^-$ described in Fig. 1; therefore,

$$\dot{N}(\omega, T, \mu) = \frac{\Psi(\omega, T, \mu)}{\hbar \omega} \int_0^h \phi(\omega) \Phi(\omega, z') dz' \tag{4}$$

where $\Phi(\omega, z') = \frac{k_0^2}{\pi^2} \text{Re}\left\{ i \int_0^\infty \varepsilon''(\omega) \left[ F_1(\omega, k_\parallel, z', 0) + F_2(\omega, k_\parallel, z', h) \right] k_\parallel dk_\parallel \right\} \tag{5}$

Here, $k_0 = \omega/c$ is the wavevector in vacuum, $k_\parallel = \sqrt{k_x^2 + k_y^2}$ is the parallel wavevector, and functions $F_1$ and $F_2$ represent the solution of multilayer Green's function for emission originated from $z'$ towards $z < 0$ and $z > h$ regions, respetively.[31,34] Note that $F_{1,2}$ depend on the dielectric function and thickness of each layer.

Applying Boltzmann approximation discussed in the SM, the local radiative recombination rate is expressed as

$$B(z') = \int_{\omega_g}^\infty B_\omega(\omega, z') d\omega = \int_{\omega_g}^\infty \frac{\phi(\omega) \exp(-\hbar \omega / k_B T)}{n_i^2} \Phi(\omega, z') d\omega \tag{6}$$

where $n_i$ is the intrinsic carrier concentration that is a function of temperature. The radiative recombination rate of the film is expressed as $B_{FE} = \frac{1}{h} \int_0^h B(z') dz'$.

For a free-standing film in the far field, the radiative recombination coefficient of the cell is simplified as

$$B_{FE} = \int_{\omega_g}^\infty \frac{\phi(\omega) \exp(-\hbar \omega / k_B T)}{n_i^2 h} \int_0^{k_0} \frac{k_\parallel dk_\parallel}{2\pi^2} \sum_{\gamma=s,p} \left(1 - |R^\gamma|^2 - |T^\gamma|^2\right) d\omega \tag{7}$$

where $R^\gamma$ and $T^\gamma$ are the reflection and transmission coefficients of the thin film when light is incident from air. The superscript $\gamma$ represents the polarization state of light ($s$ or $p$). The summation in Eq. (7) represents the absorptance of the film and may be derived directly from FE or indirectly using Kirchhoff's law.[34,35] Equation (7) provides a modified van Roosbroeck-Shockley relation that is applicable to thin films. The conventional van Roosbroeck-Shockley relation is discussed in SM.

Take a free-standing intrinsic InAs film with $E_g$ = 0.354 eV at $T$ = 300 K as an example. The radiative recombination coefficient $B_{FE}$ or $B_{ext}$ as a function of the film thickness is calculated from Eqs. (7) or (S8), respectively, for comparison. It can be seen that $B_{ext}$ based on the van Roosbroeck-Shockley relation decrease monotonically as $h$ increases. Due to interference effects, $B_{FE}$ oscillates and deviates significantly from $B_{ext}$ when $h$ < 10 μm. Note that the ratio of $B_{FE}$ and $B_{ext}$ is 4.3 and 3.6 times when $h$ = 10 nm and 20 nm, respectively. The calculated $B_{int}$ using Eq. (S7) is $2.36 \times 10^{-16}$ m$^3$/s, which is about 30 times that of $B_{ext}$ for very small $h$, where $A(\omega) \approx \alpha(\omega)h$ holds. The calculation based on fluctuational electrodynamics fully capture the wavy feature due to interference when the film thickness is less than 10 μm.

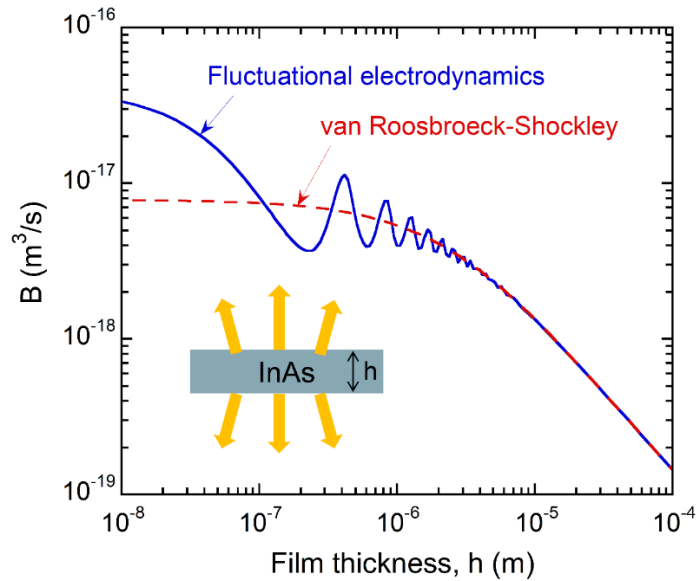

Fig. 2.　　Comparison of the radiative recombination coefficient of an intrinsic InAs cell (with varying thickness) calculated using the van Roosbroeck-Shockley relation ($B_{ext}$) and fluctuational electrodynamics ($B_{FE}$).

The effect of surrounding structure on $B(z')$ is considered in two scenarios. One is a TPV and the other is a TPV with a back gapped reflector (TPV-BGR), as shown in the inset of Fig. 3. The emitter is a bulk tungsten with a 30-nm-thick indium tin oxide (ITO) film to enhance plasmonic resonances with the 400-nm InAs cell in the near field.[6] A BGR is placed at the lower region with a fixed 10 nm vacuum gap between the InAs cell and a 100-nm-thick gold film. The vacuum gap spacing ($d$) between the emitter and the InAs cell is taken as a variable. Similar structures have been investigated previous.[36,37] The local radiative recombination coefficient expressed in Eq. (6) depends on the cell's temperature and dielection function as well as the dielectric functions and thicknesses of the surrounding materials. As discussed in the SM, for InAs with a doping level less than $10^{17}$ cm$^{-3}$, the intrinsic dielectric function may be assumed.

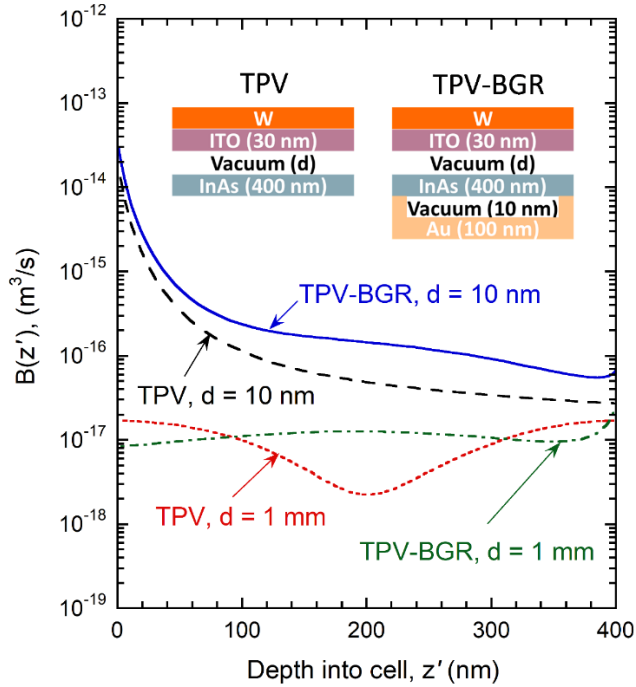

Fig. 3. Local radiative recombination coefficient for a TPV and TPV-BGR configurations, as shown in the inset, with both the near-field ($d$ = 10 nm) and far-field ($d$ = 1 mm) vacuum gaps. The active region is modeled as intrinsic InAs at 300 K.

In the near field, the radiative recombination is highly localized and structure dependent. As shown in Fig. 3, as $d$ is reduced from 1 mm to 10 nm, $B(z')$ increases by orders of magnitude, especially near the front surface due to surface plasmon polaritons (SPPs). SPPs are excited at large $k_\parallel$, resulting in very small penetration depth.[38] The frustrated modes also contribute to the enhancement with a much large penetration, resulting in the enhanced $B(z')$ for the TPV at $d = 10$ nm. The frustrated modes are associated with $k_0 < k_\parallel < m^2 k_0$, where the refractive index $m$ is between 3.5 and 3.9 for InAs in the spectral region of interest. Adding a BGR gives rise to a higher $B(z')$ near the back as well as in the middle region of the cell due to multiple reflections between the plasmonic emitter and the BGR, which affect the TPV-BGR configuration for both $d = 10$ nm and 1 mm. In the far field ($d = 1$ mm) without BGR, the profile of $B(z')$ for the TPV is nearly symmetric about the center of the cell with a valley in the middle region; this is caused by wave interference within the InAs film. Contour plots of $B(\omega, z')$ are shown in the SM to help understanding the spectral distribution. It should be noted that $B(z')$ for a free-standing InAs film (not shown) is very similar to the case with the TPV when $d = 1$ mm, dispite that the tungsten-ITO emitter is highly reflecting in the far field.

At the front surface of the cell for $d = 10$ nm, $B(z' = 0^+)$ reaches $1.8 \times 10^{-14}$ m$^3$/s for the TPV and $3.0 \times 10^{-14}$ m$^3$/s for the TPV-BGR, which are about 75 and 126 times that of $B_{\text{int}}$ predicted by the van Roosbroeck-Shockley relation. Some of the luminescently emitted photons are reabsorbed by the cell, resulting in photon recyclying. However, photon recycling is mainly due to frustrated modes and should not exceed $B_{\text{int}}$. In evaluating the photon recycling portion using near-field formulism,[30] one must set up an upper limit or cutoff $k_\parallel$ for the integration over $k_\parallel$ to be bounded.[39,40] Hence, the obtained $B(z')$ based on fluctuational electrodynamics may be assumed as the local "internal" radiative recombination coefficient near the front surface by negecting photon recycling. For the TPV-BGR configuration at $d = 10$ nm, the radiative recombination coefficient averaged over the cell, $B_{FE}$, is $6.9\times 10^{-16}$ m$^3$/s, which is nearly four times that of $B_{int}$. The BGR can significantly enhance photon recycling and improve the performance as discussed previously.[37] The associated large radiative recombination rate may also affect the dark current of the cell.[24]

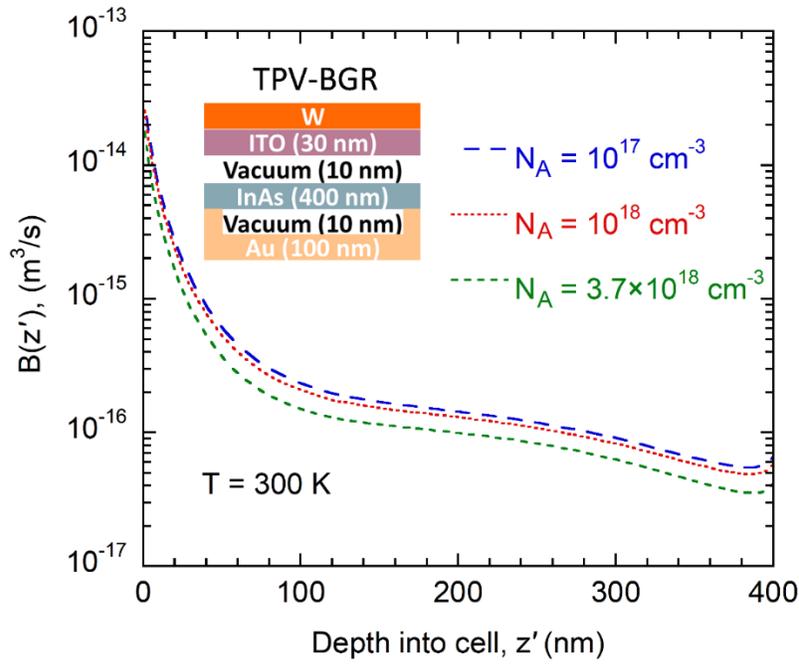

Fig. 4. Doping effect on $B(z')$ for the near-field TPV-BGR configuration. The result for $N_A$ = $1\times10^{17}$ cm$^{-3}$ is essentially the same as that for intrinsic InAs.

The doping effect on $B(z')$ for a p-type InAs film in the TPV-BGR configuration is shown in Fig. 4 at 300 K for $d = 10$ nm. The dielectric function of InAs as a function of temperature and frequency based on Milovich et al.[16] are outlined in SM. The effective density of states of the valance band $N_v = 3.7 \times 10^{18}$ cm$^{-3}$. When the donor concentration $N_A$ exceeds $N_v$, the semiconductor becomes degenerate and metallic. The spatial profile of $B(z')$ decreases with increasing doping level. The averaged $B_{FE}$ for the cell decreases from $6.9 \times 10^{-16}$ m³/s for $N_A \leq 1 \times 10^{17}$ cm$^{-3}$ to $4.1 \times 10^{-16}$ m³/s for $N_A = 3.7 \times 10^{18}$ cm$^{-3}$. As shown in Fig. S1(c), doping also affects the luminescent coefficient, resulting in a large portion of the photons emitted above the bandgap being due to free carriers.

A near-field TR cell is considered by reversing the previous TPV-BGR configuration as illustrated in the inset of Fig. 5(a) to further demonstrate the significance of the above-bandgap thermal radiation and luminescent coefficient. The side of InAs film with a BGR structure is fixed at 600 K, and the side of tungsten and ITO is fixed at 300 K. The dielectric function of the cell is modeled using $p$-type InAs with a donor concentration equal to $N_v$ at 600 K. As summarized in the SM, detailed balance analysis is used to calculate the performance.[2,24,41] The luminescent emission and thermal emission spectra for a bias voltage $V = -0.052$ V are shown in Fig. 5(a), considering free-carrier contribution as thermal emission. This bias gives a negative photon chemical potential $\mu = eV = -0.052$ eV, corresponds to the maximum power condition of this TR device. The ideal assumption with $\phi = 1$ treats all the above-bandgap absorption as due to interband transition and overpredicts the luminescent emission as shown in Fig. 5(a). At a given frequency, thermal emission is scaled to $\varepsilon''_{fc}(\omega)\Theta(\omega,T)$. The difference between the ideal and acutal luminescent emission is scaled to $\varepsilon''_{fc}\Psi(\omega,T,\mu)$, which is smaller than $\varepsilon''_{fc}\Theta(\omega,T)$ when $\mu < 0$ as in this case.

The total net luminescent emission calculated with the ideal $\phi$ assumption is 7.4 kW/m², which is 11% higher than that with the actual $\phi$. This would cause more than 10% overprediction of both the maximum power density and the efficiency of the TR as shown in Fig. 5(b). Though the calculations are oversimplified and the structure used is not optimized for a TR device, the results provide evidence of the effect of thermal emission with high dopant concentrations. Hence, the spatial effect and doping effects must be considered for thin-film near-field photonic energy converter.

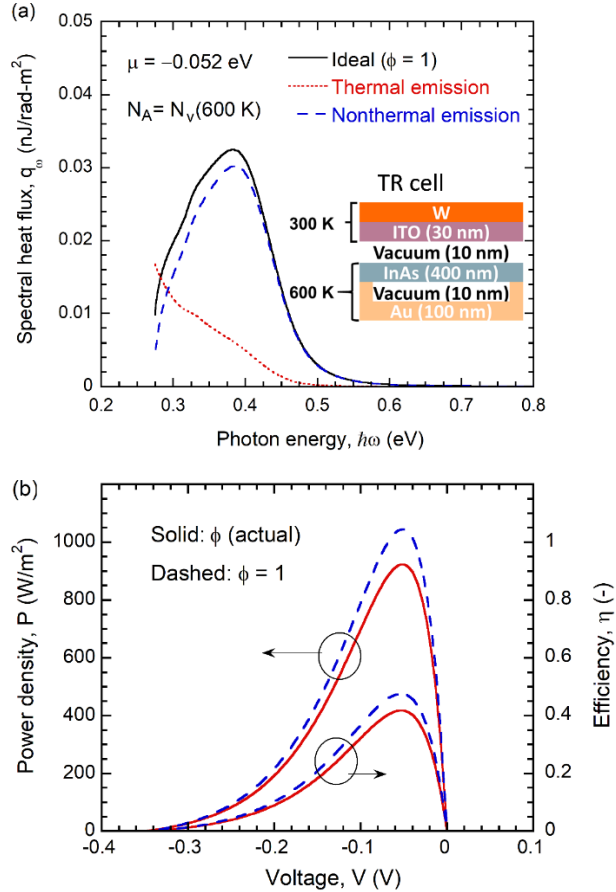

Fig. 5. (a) The net spectral heat flux due to thermal and nonthermal emission for a TR cell configuration shown in the inset for $\mu = -0.052$ eV. The active region consists a *p-n* junction or diode, though, the dielectric function is treated as uniform with *p*-doped InAs with $N_\text{A} = N_\text{v} @ 600 \text{ K} = 1.05 \times 10^{19}$ cm$^{-3}$. The net spectral heat flux by assuming ideal luminescence coefficient is also shown for comparison. (b) Power density and efficiency as functions of the bias voltage for the actual and ideal luminescence coefficient.

In summary, this work defines and develops a formulism for calculating the local radiative recombination coefficient based on fluctuational electrodynamics that is applicable to both the

near- and far-field photonic energy converters. SPPs can modify the radiative recombination coefficient near the surface by several orders of magnitudes. The spatial profile of radiative recombination coefficient can be tuned by modifying the surrounding structure and materials. The use of a luminescence coefficient allows the distinction of the luminescent emission from that of thermal emission due to the above-bandgap free-carrier contribution. It also enables more accurate modeling of photonic energy converters that employ heavily doped semiconductors.

**Acknowledgements**

D.F. and Z.M.Z. would like to thank the support from the U.S. Department of Energy (DOE), Office of Science, Basic Energy Sciences (DE-SC0018369). D.F. would also like to thank Dr. J.-J. Greffet for helpful discussions.

# Supplementary Material for

# Spatial and doping effects on radiative recombination in thin-film near-field photonic energy converters

Dudong Feng, Shannon K. Yee, and Zhuomin M. Zhang*

George W. Woodruff School of Mechanical Engineering,
Georgia Institute of Technology, Atlanta, GA 30332, USA

*Corresponding author: zhuomin.zhang@me.gatech.edu

## S1. Derivation of the local radiative recombination coefficient

The (spectral) radiative recombination coefficient is used to calculate the external luminescence (towards both the upper and lower regions) of a cell of thickness $h$ as follows:

$$\dot{N}(\omega, T, \mu) = h B_\omega(\omega) n p \tag{S1}$$

where $n$ and $p$ are the electron and hole concentrations, respectively.[1,2] Oftentimes, the emission at thermal equilibrium ($\mu = 0$) is subtracted so that

$$B_\omega(\omega) = \frac{\dot{N}(\omega, T, \mu) - \dot{N}(\omega, T, 0)}{h(np - n_i^2)} \tag{S2}$$

The Boltzmann approximation of the modified Bose-Einstein distribution is expressed as

$$\Psi_B(\omega, T, \mu) = \exp(\mu/k_B T)\exp(-\hbar\omega/k_B T) \tag{S3}$$

Furthermore, the product of the electron and hole concentrations may be expressed as[1,3]

$$np = n_i^2 \exp(\mu/k_B T) \tag{S4}$$

Equation (S2) may be simplified using the expressions of Eqs. (4), (S3), and (S4) to give

$$B_\omega(\omega) = \frac{1}{h}\int_0^h B_\omega(\omega, z')dz' = \frac{\exp(-\hbar\omega/k_B T)}{n_i^2 h}\int_0^h \phi(\omega)\Phi(\omega, z')d\omega \tag{S5}$$

Hence, the local, spectral radiative recombination coefficient becomes

$$B_\omega(\omega, z') = \frac{\phi(\omega)\exp(-\hbar\omega/k_B T)}{n_i^2}\Phi(\omega, z') \tag{S6}$$

It may be integrated over the frequency as shown in Eq. (6). The application condition of Boltzmann approximation depends on the injected carrier density and doping level of the semiconductor materials. The error of Boltzmann approximation has been discussed in Feng et al.[4]

Rigorously speaking, *B* defined above is the "external" radiative recombination coefficient because it is related to external luminescene. However, the word "external" is omitted for brievity, unless clarifications are necessary. Portion of the emitted photons are reabsorbed by the cell and the reabsorption process is called photon recycling[5,6]. Hence, the "internal" radiative recombination coefficient should be greater than the radiative recombination coefficient given previously. Since *B* is related to electroluminescence and recombination lifetime, it is an important parameter in radiative energy converters.[7]

## S2. Expressions of the van Roosbroeck-Shockley relation

The van Roosbroeck-Shockley relation provides an expression of the internal radiative recombination coefficient with the absorption coefficient $\alpha(\omega)$ as follows:[8]

$$B_{\text{int}} = \int_{\omega_g}^{\infty} \frac{k_0^2 m^2}{\pi^2 n_i^2} \alpha(\omega) \exp\left(-\frac{\hbar\omega}{k_B T}\right) d\omega \tag{S7}$$

where *m* is the refractive index of the cell material. It is related to the volumetric radiative recombination rate by $R_{\text{rad}} = np B_{\text{int}}$. The definition given in Eq. (S7) may be thought as an intrinsic or bulk radiative recombination coefficient since it does not involve the boundaries or anything outside the medium. Trupke et al.[1] related the absorption coefficient with the photoluminescence spectrum to calculate the radiative recombination coefficient. Following their derivation, one obtains a corresponding "external" radiative recombination coefficient of a layer as follows.

$$B_{\text{ext}} = \int_{\omega_g}^{\infty} \frac{k_0^2}{2\pi^2 n_i^2 h} A(\omega) \exp\left(-\frac{\hbar\omega}{k_B T}\right) d\omega \tag{S8}$$

where $A(\omega)$ is the spectral absorptance of the cell layer considering multiple reflections at the interfaces. The absorptance is the same for incidence from either above or below the cell when it is a free-standing film. Here, the external luminescent emission considers both the upper and lower hemisphere. While multiple reflections are included in calculating $A(\omega)$, the original formulation did not consider interference effects, which are important for thin films whose thickness is on the order or smaller than the wavelength of interest.

When the cell thickness $h$ is small and, $A(\omega) \approx \alpha(\omega)h$ for near-normal incidence or emission. The relationship, $B_{\text{int}} = 2n^2 B_{\text{ext}}$, between internal and external recombination can be drawn by comparison of Eqs. (S7) and (S8). The factor of $2n^2$ was obtained by Yablonovitch[9] based on geometric or ray optics. Considering hemispherical emission/absorption, the factor $2n^2$ still holds for large enough $n$ that makes the refraction angle in the cell to be sufficiently small. When $n \to 1$ (the refractive index of the cell is about the same as the surroundings), surface reflection is negligibly small. Furthermore, integration of the path length $h/\cos\theta$ (where $\theta$ is the polar angle) over the hemisphere gives $A(\omega) \approx 2\alpha(\omega)h$ for very small $h$. The factor of 2 is due to the angular effect that gives rise to a longer path length inside the cell layer. Substiting $A(\omega) \approx 2\alpha(\omega)h$ into Eq. (S8) gives the same result as Eq. (S7), that is, $B_{\text{int}} = B_{\text{ext}}$.

It should be noted that Eq. (S8) is appropriate for most practical applications such as solar cells. Nevertheless, it will produce a large relatively error as the cell thickness is reduced to submicron regime as illustrated in Fig. 2. In the near-field regime, the radiative recombination processes also depend on the surrounding materials and geometric parameters such as the vacuum gap distance as discussed in the main text.[3,4,6,7]

## S3. Dielectric function model of doped InAs

The dielectric function of InAs is depicted by the summation of absorption due to interband transitions, lattice resonance (optical phonon), and free-carrier absorption. Milovich *et al.*[10] developed a comprehensive model for the dielectric function of InAs as a function of dopant concentration and temperature. Since the absorption edge of a narrow bandgap semiconductor can be affected under the heavily doped condition, the Moss-Burstein effect (or shift) needs to be considered.[11,12] A brief summary of the dielectric function model is given in the following.

A Drude-Lorentz model is used to calculate the contribution by phonons (lattice vibration) and free carriers to the frequency-dependent dielectric function as[13]

$$\varepsilon_{la} + \varepsilon_{fc} = \varepsilon_\infty \left[ \frac{\omega_{LO}^2 - \omega_{TO}^2}{\omega_{TO}^2 - \omega^2 - i\omega\gamma} - \frac{\omega_p^2}{\omega(\omega + i\Gamma)} \right] \quad (S9)$$

where $\varepsilon_\infty$ is the high-frequency constant, $\omega_{LO}$ and $\omega_{TO}$ are the longitudinal and transverse optical phonon frequencies, $\gamma$ and $\Gamma$ are the damping rates for phonons and free carriers, respectively, and $\omega_p$ is the plasma frequency divided by $\varepsilon_\infty^{1/2}$.

The dielectric function due to interband transitions is calculated based on the interband absorption coefficient, where the Moss-Burstein shift is considered for heavily-doped narrow bandgap semiconductors. The Fermi energy level is solved as a function of dopant concentration using the relation given in Sijerčić et al.[14] with the InAs parameters. The interband absorption coefficient $\alpha_{ib}(\omega)$ as a function of frequency and dopant concentration is then evaluated using Eq. (16) or (20) in Ref. [11]. A Kramers-Kronig transformation of $\kappa_{ib}(\omega) = \alpha_{ib}(\omega)/(2k_0)$ gives the contribution of the interband transition to the refractive index as $m_{ib}(\omega)$,[10] which is negligibly small except at frequencies near the band edge. Here, $\kappa_{ib}(\omega)$ is the absorption index or imaginary part of the complex refractive index due to interband transitions. The dielectric function associated with interband transition is then expressed as

$$\varepsilon_{ib} = \left[ \sqrt{\varepsilon_\infty} + m_{ib} + i\kappa_{ib} \right]^2 \quad (S10)$$

The combined frequency-dependent complex dielectric function of InAs is calculated by the summation $\varepsilon = \varepsilon_{ib} + \varepsilon_{la} + \varepsilon_{fc}$. At $\omega < \omega_g$, $\varepsilon = \varepsilon_\infty + \varepsilon_{la} + \varepsilon_{fc}$ as expected. At $\omega > \omega_g$, the phonon contribution $\varepsilon_{la}$ becomes negligibly small, resulting in the following approximation:

$$\varepsilon \approx m^2 + i\left[ \varepsilon''_{ib} + \varepsilon''_{fc} \right] \quad (S11)$$

where $m$ is the real part of the refractive index, and $\varepsilon''_{ib} + \varepsilon''_{fc} = \varepsilon''$ is the imaginary part of the dielectric function. In general, at high frequencies, $\varepsilon'' \ll m^2$.

The imaginary part of the dielectric function may be related to the coresponding absorption coefficient by[13]

$$\varepsilon''_{ib} + \varepsilon''_{fc} = m(\alpha_{ib} + \alpha_{fc})/k_0 \quad (S12)$$

Hence, the luminescence coefficient given in Eq. (3) may be written in terms of the absorption coefficient as $\phi = \alpha_{ib}/(\alpha_{ib} + \alpha_{fc}) = \alpha_{ib}/\alpha$, which was a factor used to calculate the quantum efficiency of InSb photodiodes.[12]

## S4. Calculated spectral dielectric function and luminescence coefficient

The doping level and temperature effects on the dielectric function and luminescence coefficient of a $p$-type InAs is shown in Fig. S1. The absorption coefficient is proportional to the imaginary part of the dielectric function according to Eq. (S12). Hence, a large $\varepsilon''$ As shown in Figs. S1(a) and S1(b), a large spike around 0.027 eV represents the absorption due to optical phonon. As the doping level increases, the broadband contribution of free electron also increases.

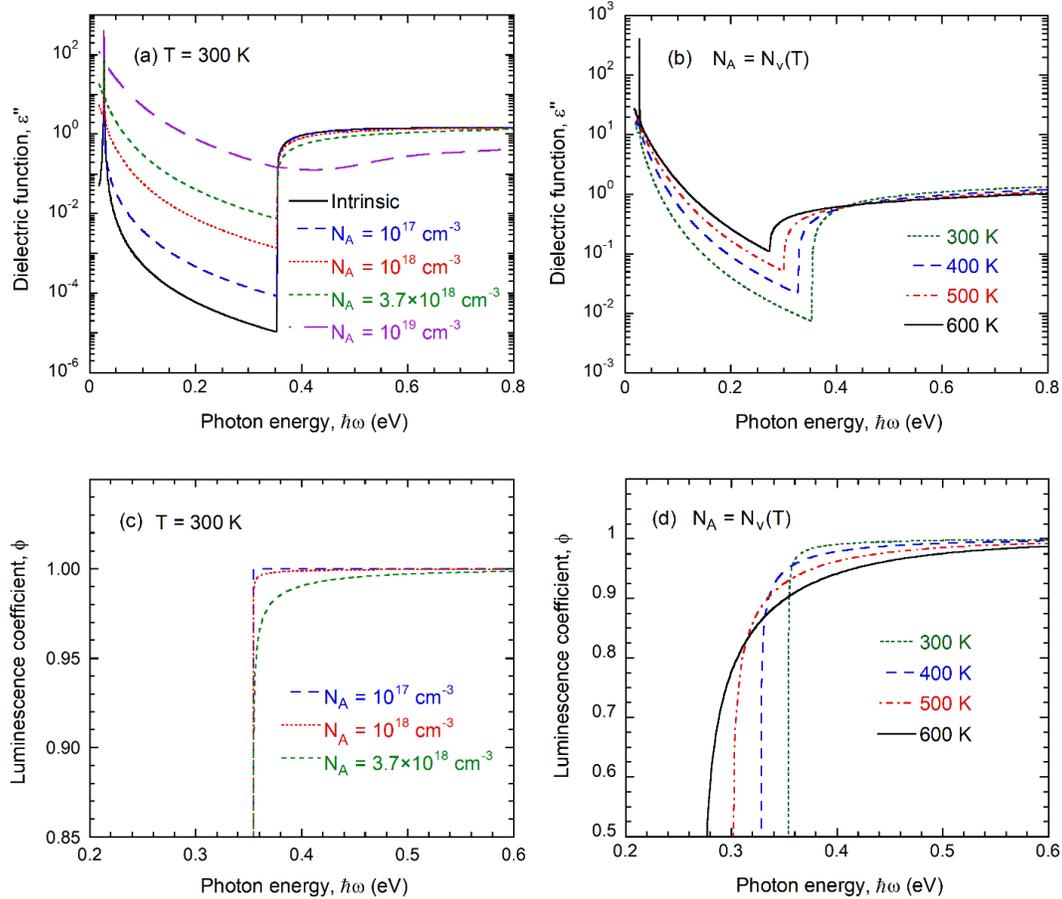

Fig. S1. The imaginary part of dielectric function of $p$-InAs: (a) for different doping levels at 300 K, and (b) at different temperatures when the acceptor concentration is set to equal to effective density of states of the valence band. (c) and (d) The luminescence coefficient spectra corresponding to (a) and (b), respectively.

Interband absorption happens above the bandgap energy (0.354 eV at 300 K), where $\varepsilon''$ sharply jump up. The effective density of states of the valence band at 300 K is $N_v = 3.7 \times 10^{18}$ cm$^{-3}$. When the donor concentration $N_A$ exceeds $N_v$, the Fermi level will lie below top of the valance band and the semiconductor becomes degenerate.[15] When the donor concentration greatly exceeds $N_v$, the bandgap feature is submerged by excess free carriers. As shown in Fig. S1(a) when $N_A = 10^{19}$ cm$^{-3}$, degenerate InAs become metallic and gapless, although $\varepsilon''$ is still much smaller than a typical metal in the mid-infrared region.

The bandgap energy is a strong function of temperature and shifts toward lower energies as temperature increases. Since $N_v$ is proportional to $T^{1.5}$,[15] the $N_A$ value used for Fig. S1(b) also increases with temperature, giving $N_A = 1.05 \times 10^{19}$ cm$^{-3}$ at 600 K. Therefore, the increase of free-carrier absorption with temperature shown in Fig. S1(b) is due to both the temperature-dependent damping rate and plasma frequency.

The corresponding luminescence coefficient spectra are plotted in Figs. S1(c) and S1(d), respectively. The doping effect on $\phi$ is negligibly small towards higher frequencies. At 300 K, $\phi$ decreases somewhat as the dopant concentration increases up to the degenerate level. Even with $N_A = 3.7 \times 10^{18}$ cm$^{-3}$, $\phi > 0.95$ at $\hbar\omega > 0.357$ eV. With increasing temperature and concentration, $\phi$ becomes smaller as shown in Fig. S1(d). Hence, the assumption of an ideal luminescent coefficient ($\phi = 1$) may give rise to a relatively large error in the prediction of thermoradiative device performance because the cell is operating at higher temperatures.

## S5. Local spectral radiative recombination coefficient

The contour plots of $B(\omega, z')$ for the corresponding curves of $B(z')$ shown in Fig. 3 are displayed in Fig. S2 for the TPV and TPV-BGR structures for vacuum gap $d = 10$ nm and 1 mm. To illustrate the contour clearly, Figs. S2(a) and S2(b) are plotted in log scale while Figs. S2(c) and S2(d) are plotted in linear scale for $B(\omega, z')$. The large enhancement of surface plasmon polaritons (SPP) on $B(\omega, z')$ for $z'$ close to zero is evident and occurs in a broad spectral region above the bandgap. In the far-field cases, $B(\omega, z')$ is much smaller and mainly in the region for $0.354$ eV $< \hbar\omega < 0.42$ eV. For the TPV without BGR, the shape is nearly symmetric about the middle of the cell, as shown in Fig. S2(c). On the other hand, it can be seen from Fig. S2(d), there appears to be some SPP effect on the back side near the Au reflector, resulting the asymmetry function of $B(z')$. However, the enhancement brought by SPPs on the back side is much smaller than that brought by SPPs due to the W-ITO structure.

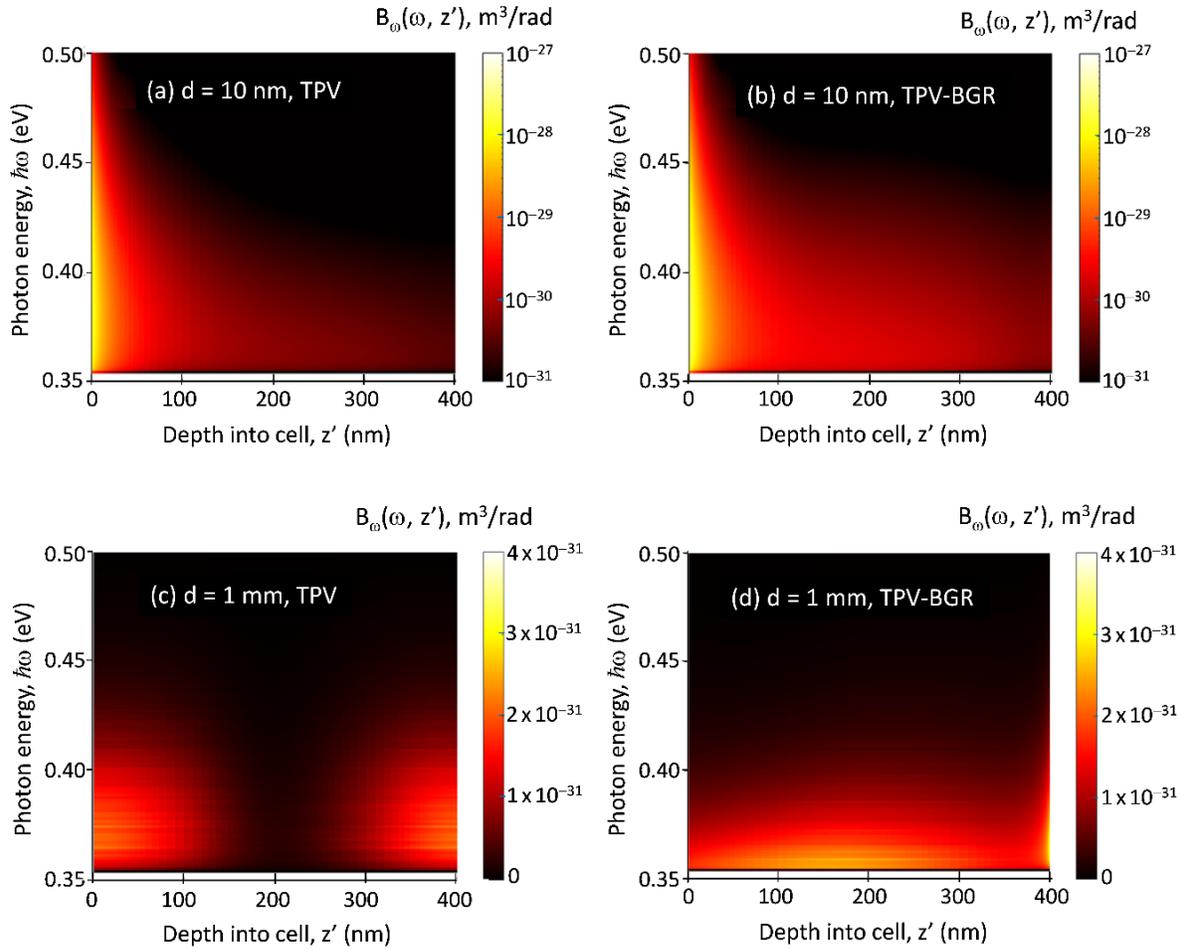

Fig. S2. Counter plots of the local spectral radiative recombination coefficient $B(\omega, z')$: (a) TPV configuration with $d$ = 10 nm; (c) TPV-BGR configuration with $d$ = 10 nm; (c) TPV configuration with $d$ = 1 mm; (b) TPV-BGR configuration with $d$ = 1 mm.

## 6. Thermoradiative cell

A thermoradiative (TR) device operates when the photovoltaic (PV) cell is at a temperature higher than the surroundings. The *p-n* junction is negatively biased so that it emits fewer photons than the corresponding case at thermodynamic equilibrium. There is an associated electron-hole pair generation to recover the equilibrium population, resulting in photocurrent and power output to a load. Detailed descriptions and derivations can be found from the literature.[16,17] By neglecting all nonradiative recombination process, the detailed balance model is briefly summarized below based on the configuration shown as the inset of Fig. 5(a).

The total net heat transfer per unit area ($q_{net}$) out of the cell may be obtained by integrating Eq. (1) over all the frequencies with a modification to the function $W$ as follows:

$$W_{mod}(\omega, z') = \phi(\omega)\varepsilon''(\omega)\left[\Psi(\omega, T_c, \mu) - \Theta(\omega, T_s)\right] \\ + \left[1 - \phi(\omega)\right]\varepsilon''(\omega)\left[\Theta(\omega, T_c) - \Theta(\omega, T_s)\right] \tag{S13}$$

where $T_c$ and $T_s$ are the temperature of the cell and that of the colder region, respectively.

The current density is written as

$$J = e\int_{\omega_g}^{\infty} \dot{N}_1(\omega, T_c, T_s, \mu) d\omega \tag{S14}$$

where e is the elementary charge, and $\dot{N}_1(\omega, T_c, T_s, \mu)$ is the net photon flux to the front side expressed as

$$\dot{N}_1(\omega, T_c, T_s, \mu) = \frac{\Psi(\omega, T_c, \mu) - \Theta(\omega, T_s)}{\hbar\omega}\int_0^h \phi(\omega)\Phi_1(\omega, z')dz' \tag{S15}$$

Here,

$$\Phi_1(\omega, z') = \frac{k_0^2}{\pi^2}\text{Re}\left[i\int_0^{\infty} \varepsilon''(\omega)F_1(\omega, k_{\parallel}, z', 0)k_{\parallel}dk_{\parallel}\right] \tag{S16}$$

since only the emission to the front side needs to be considered. Under the detailed balance assumption, $\mu = eV$, where $V$ is the voltage bias. The output power density and efficiency are calcualted from

$$P = |JV| \tag{S17}$$

and

$$\eta = P/q_{net} \tag{S18}$$

Note that in the calculation, the photon chemical potential is set as a constant across the cell and the dielectric function of the cell materials is assumed to be the save. However, the calculation of $\Psi(\omega, T_c, \mu)$ is based on the modified Bose-Einstein function without using the Boltzmann approximation.